\documentclass[aps,twocolumn,superscriptaddress,notitlepage,nofootinbib]{revtex4-1}

\usepackage{hyperref}
\usepackage{epsfig}
\usepackage{multirow}
\usepackage{slashed}
\usepackage{amsmath}
\usepackage{physics}
\usepackage{qcircuit}
\usepackage{braket}
\usepackage{graphicx}
\usepackage{subfigure}
\usepackage{color}
\usepackage{epstopdf}
\usepackage[inline]{enumitem}
\usepackage{bm}

\allowdisplaybreaks

\begin{document}

\title{Scattering amplitude from quantum computing with reduction formula}

\date{\today}

\author{Tianyin Li}
\affiliation{Key Laboratory of Atomic and Subatomic Structure and Quantum Control (MOE), Guangdong Basic Research Center of Excellence for Structure and Fundamental Interactions of Matter, Institute of Quantum Matter, South China Normal University, Guangzhou 510006, China}
\affiliation{Guangdong-Hong Kong Joint Laboratory of Quantum Matter, Guangdong Provincial Key Laboratory of Nuclear Science, Southern Nuclear Science Computing Center, South China Normal University, Guangzhou 510006, China }

\author{Wai Kin Lai}
\email{Corresponding author.\\
wklai@m.scnu.edu.cn}
\affiliation{Key Laboratory of Atomic and Subatomic Structure and Quantum Control (MOE), Guangdong Basic Research Center of Excellence for Structure and Fundamental Interactions of Matter, Institute of Quantum Matter, South China Normal University, Guangzhou 510006, China}
\affiliation{Guangdong-Hong Kong Joint Laboratory of Quantum Matter, Guangdong Provincial Key Laboratory of Nuclear Science, Southern Nuclear Science Computing Center, South China Normal University, Guangzhou 510006, China }
\affiliation{Department of Physics and Astronomy, University of California, Los Angeles, California 90095, USA}

\author{Enke Wang}
\email{wangek@scnu.edu.cn}
\affiliation{Key Laboratory of Atomic and Subatomic Structure and Quantum Control (MOE), Guangdong Basic Research Center of Excellence for Structure and Fundamental Interactions of Matter, Institute of Quantum Matter, South China Normal University, Guangzhou 510006, China}
\affiliation{Guangdong-Hong Kong Joint Laboratory of Quantum Matter, Guangdong Provincial Key Laboratory of Nuclear Science, Southern Nuclear Science Computing Center, South China Normal University, Guangzhou 510006, China }

\author{Hongxi Xing}
\email{Corresponding author.\\
hxing@m.scnu.edu.cn}
\affiliation{Key Laboratory of Atomic and Subatomic Structure and Quantum Control (MOE), Guangdong Basic Research Center of Excellence for Structure and Fundamental Interactions of Matter, Institute of Quantum Matter, South China Normal University, Guangzhou 510006, China}
\affiliation{Guangdong-Hong Kong Joint Laboratory of Quantum Matter, Guangdong Provincial Key Laboratory of Nuclear Science, Southern Nuclear Science Computing Center, South China Normal University, Guangzhou 510006, China }
\affiliation{Southern Center for Nuclear-Science Theory (SCNT), Institute of Modern Physics, Chinese Academy of Sciences, Huizhou 516000, China}  

\collaboration{QuNu Collaboration}

\begin{abstract}
Utilizing the Lehmann-Symanzik-Zimmermann reduction formula, we present a new general framework for computing scattering amplitudes in quantum field theory with quantum computers in a fully nonperturbative way. In this framework, one only has to construct one-particle states of zero momentum, and 
no wave packets of incoming particles are needed. The framework is able to incorporate scatterings of bound states, and is
ideal for scatterings involving a small number of particles. We expect this framework to have particular advantages when applied to
 exclusive hadron scatterings. As a proof of concept, by simulations on classical hardware, we demonstrate that in the one-flavor Gross-Neveu model, the fermion propagator,
the connected fermion four-point function, and the propagator of a fermion-antifermion bound state obtained
from our proposed quantum algorithm have the desired pole structure crucial to the implementation of the Lehmann-Symanzik-Zimmermann reduction formula.

\end{abstract}

\maketitle

\section{Introduction} 
The calculation of scattering amplitudes in quantum field theory has long been a core topic in theoretical 
particle physics~\cite{Feynman:1949zx,Eden:1966dnq,Bern:1994fz,Britto:2005fq,Bern:2007dw}. All tests of theories against experiments in particle accelerators entail theoretical predictions
of scattering amplitudes. Despite the huge success of the perturbative approach to the calculation of scattering 
amplitudes~\cite{ALEPH:2005ab,Anastasiou:2016cez,ATLAS:2019zci},
there are still circumstances in which the perturbative framework does not work, namely the cases where the coupling constants
are large, as is the case for quantum chromodynamics at low energies for instance.
To date, first-principle nonperturbative calculations of scattering amplitudes in quantum field theory are not available. The main obstacle is that real-time
dynamics cannot be simulated in traditional path-integral lattice quantum field theory~\cite{Alexandru:2016gsd}, while simulating real-time Hamiltonian evolutions in quantum field theory  
requires unbearable computational cost on a classical computer. In a series of papers, Jordan, Lee, and Preskill (JLP) proposed that, with the help of quantum
computers, simulations of Hamiltonian evolutions of scattering processes in quantum field theory can be achieved with affordable computational cost on the lattice,
making nonperturbative evaluations of scattering amplitudes possible~\cite{Jordan:2012xnu,Jordan:2011ci,Jordan:2014tma}. These works have spurred a series of research on the applications of quantum computing in
particle physics~\cite{Bauer:2022hpo}, ranging from time evolutions in quantum field
theory~\cite{Martinez:2016yna,Hu:2019hrf,Bauer:2019qxa,DeJong:2020riy,Zhou:2021kdl,deJong:2021wsd,Bepari:2021kwv,Atas:2022dqm,Yao:2022eqm} 
to calculations of nonperturbative quantities
~\cite{Lu:2018pjk,Lamm:2019uyc,Mueller:2019qqj,Roggero:2019myu,Echevarria:2020wct,Kreshchuk:2020kcz,Bauer:2021gup,Atas:2021ext,Li:2021kcs,Li:2022lyt,Gallimore:2022hai} and 
 thermodynamics at finite chemical potential~\cite{Czajka:2021yll,Czajka:2022plx,Xie:2022jgj}. 

Although the quantum-computational framework developed by JLP is fully general, it may encounter difficulties in practice.
One major difficulty is that one has to prepare spatially well-separated wave packets of incoming particles in the initial state. 
The central values of the 4-momenta of these wave packets $p_i$, as well as their Lorentz-invariant products $p_i\cdot p_j$,
set a constraint 
\begin{align}
a\ll 1/|p_i^\mu|,1/\sqrt{|p_i\cdot p_j|}\ll L \,,\label{eq:constraint_1}
\end{align}
on the lattice spacing $a$ and the lattice size $L$; while the spatial separation distances
$d_{ij}$ between the initial-state wave packets set a constraint 
\begin{align}
L\gg d_{ij} \,, \label{eq:constraint_2}
\end{align}
on the lattice size. In addition, the separations $d_{ij}$ have to be wide, meaning that $d_{ij}\gg 1/\Delta p_i^\mu$, where $\Delta p_i^\mu$ is the
uncertainty of the wave packet of the $i$th incoming particle in momentum space, and we require $\Delta p_i^\mu\ll|p_i^\mu|$ so that the 
wave packets are narrow enough to mimic a scattering process of definite incoming momenta. Constraint Eq.~(\ref{eq:constraint_1}) is 
required for reliable simulations of incoming particles with definite 4-momenta on the lattice, and can be potentially improved
by introducing factorization theorems using the method of effective field theory~\cite{Bauer:2021gup}. Constraint Eq.~(\ref{eq:constraint_2}) is due to the 
introduction of wave packets, which generally implies a larger lattice size than required by Eq.~(\ref{eq:constraint_1}). 
Another feature of the JLP formalism is that the wave packets are first prepared with the coupling constant turned off. Therefore, the incoming particles cannot be bound states.
The coupling constant is subsequently adiabatically turned on before the scattering occurs and adiabatically turned off after the scattering occurs.  
To ensure adiabaticity, a long time span of evolution is required. In addition, one has to insert backward evolutions
in order to eliminate unwanted broadening of wave packets during the adiabatic turn on and turn off of the coupling constant, thus increasing the time complexity. 
In fact, in the strong coupling regime, most theoretical uncertainties come from the adiabatic turn on and turn off of the coupling constant~\cite{Jordan:2012xnu}.

We note that, in the conventional perturbative approach, scattering amplitudes are computed using the Lehmann-Symanzik-Zimmermann (LSZ) reduction formula~\cite{LSZ}, which relates
scattering amplitudes to $n$-point correlation functions, which in turn can be expanded as power series in the coupling constant using the Feynman diagram technique. The LSZ reduction formula, being a nonperturbative relation, is a natural
alternative starting point for the evaluation of scattering amplitudes with quantum computers in a fully nonperturbative way. 
In this work, we propose a quantum computational framework for calculating scattering amplitudes using the LSZ reduction formula.
In this framework, in order to evaluate scattering amplitudes, one calculate $n$-point correlation functions on a quantum computer.
We will see that,  this approach is ideal for scattering processes involving a small number of particles,
and will have potential applications in exclusive strong-interaction processes such as $2\to 2$ scatterings of pions or nucleons.

In the following, we first give a short review of the LSZ reduction formula. We then propose a quantum algorithm which utilizes the LSZ reduction formula to compute scattering amplitudes, and discuss its features and advantages. 
After that, as a proof of concept, in a simple model, the one-flavor Gross-Neveu model, we simulate the fermion propagator, the connected fermion
four-point function, and the propagator of a fermion-antifermion bound state with our proposed quantum algorithm on classical hardware.
We give a conclusion at the end.

\section{LSZ reduction formula}
The LSZ reduction formula relates the scattering amplitude of a given scattering process to correlation functions of fields in the vacuum~\cite{LSZ}. 
For instance, consider the scattering process $h(\bm{k}_1)+\dots+h(\bm{k}_{n_{\rm in}})\to h(\bm{p}_1)+\dots+h(\bm{p}_{n_{\rm out}})$, where $h$ is some spin-$0$ particle with mass $m$
annihilated by a scalar field $\phi$.
Using the LSZ reduction formula, the scattering amplitude $\mathcal{M}$ can be written as
\begin{align}
&i\mathcal{M}=R^{n/2}
\lim_{\begin{array}{c}p_i^2\to m^2\\k_j^2\to m^2\end{array}}G(\{p_i\},\{k_j\})\nonumber\\
&  \quad\quad\times\left(\prod_{r=1}^{n_{\rm out}} K^{-1}(p_r)\right)
\left(\prod_{s=1}^{n_{\rm in}} K^{-1}(k_s)\right)\,,\label{eq:LSZ}
\end{align}
where $n=n_{\rm in}+n_{\rm out}$.
The $G(\{p_i\},\{k_j\})$ is the connected $n$-point function in momentum space, given by
\begin{align}
&\,\,\,\,\,\,\,G(\{p_i\},\{k_j\})\nonumber\\
&=
\left( \prod_{i=1}^{n_{\rm out}} \int d^4x_i\, e^{i p_i\cdot x_i}\right)\left(\prod_{j=1}^{n_{\rm in}-1} \int  d^4y_j\, e^{-i k_j\cdot y_j}\right)\nonumber\\
& \times
\langle \Omega|T\left\{\phi(x_1)\cdots\phi(x_{n_{\rm out}})\phi^\dagger(y_1)\cdots \phi^\dagger(y_{n_{\rm in}-1})\phi^\dagger(0)\right\}|\Omega\rangle_{\rm con}\,,
\label{eq:G_m_n}
\end{align}
where $T$ denotes time ordering, the subscript ``con" denotes the connected part, and $|\Omega\rangle$ is the interacting vacuum, i.e. the ground state.
The $K(p)$ is the two-point function in momentum space, also called the propagator, given by
\begin{align}
K(p)
&=\int d^4x\,e^{ip\cdot x} \langle \Omega|T\{\phi(x) \phi^\dagger(0)\}|\Omega\rangle_{\rm con}\,.\label{eq:G}
\end{align}
The factor $R$ is the field normalization, defined by
\begin{align}
R&=|\langle \Omega|\phi(0)|h(\bm{p}=0)\rangle|^2\,, \label{eq:R}
\end{align}
where $|h(\bm{p}=0)\rangle$ denotes the state with a single particle $h$ with zero spatial momentum.
The generalization of Eq.~(\ref{eq:LSZ}) to cases which involve multiple types of massive particles with arbitrary spin is trivial,
with suitable inclusions of polarization tensors and spinors on the right-hand side of Eq.~(\ref{eq:LSZ}).
In essence, the LSZ reduction formula Eq.~(\ref{eq:LSZ}) says that the scattering amplitude is simply a connected $n$-point function in
momentum space with momenta put on-shell, with external-leg propagators amputated. The field normalization factors $\sqrt{R}$
on the right-hand side of Eq.~(\ref{eq:LSZ}) ensure that the scattering amplitude, as a physical observable, is independent of the 
normalization of the field operators which create or annihilate the external particles. It should be noted that the connected $n$-point function
$G(\{p_i\},\{k_j\})$ has simple poles at $p_i^2,k_j^2= m^2$, and so is divergent when the momenta are put on shell. On the other hand,
the propagator $K(p)$ also has a simple pole at $p^2=m^2$, namely
\begin{align}
K(p)\overset{p^2\to m^2}{\longrightarrow}\frac{iR}{p^2-m^2+i\epsilon}\,.\label{eq:2_pole}
\end{align}
Therefore, in Eq.~(\ref{eq:LSZ}), the pole singularities in $G(\{p_i\},\{k_j\})$ cancel with those in the $K(p)$ factors, giving a finite scattering
amplitude. In practice, when the continuum theory is approximated by a theory on the lattice, these singularities are tamed and the 
pole structure $\frac{1}{p^2-m^2+i\epsilon}$ is approximated by some bounded function of $p^2$ that approaches it in the continuum and infinite-volume limits.
A study of finite-volume effects on Minkowski correlation functions has been done in 
Ref.~\cite{Briceno:2020rar}.

According to Eq.~(\ref{eq:LSZ}), the computation of the scattering amplitude is broken down to the computation of three objects: the connected $n$-point
 function $G(\{p_i\},\{k_j\})$, the propagator $K(p)$, and the field normalization $R$. 
Implementing the computation of these objects on a quantum computer will involve three steps: (1) the spatial dimensions are discretized into a lattice,
(2) the field degrees of freedom are mapped to qubits, (3) a suitable quantum algorithm is constructed to evaluate the three objects individually. 
For gauge theories, step (1) can be achieved in the standard way under the Kogut-Susskind Hamiltonian formalism~\cite{Kogut:1974ag,Kogut:1982ds}, and alternative approaches have been proposed ~\cite{Anishetty:2009nh,Raychowdhury:2019iki,Buser:2020cvn}. Step (2)
can be done straightforwardly for fermionic degrees of freedom~\cite{Jordan_Wigner,Bravyi,Backens:2019}, while for bosonic degrees of freedom and in particular gauge bosons considerable
progress has been made~\cite{Klco:2018zqz,Byrnes:2005qx,Zhang:2018ufj,Unmuth-Yockey:2018xak,Raychowdhury:2019iki,Buser:2020cvn,Alexandru:2019nsa,Ji:2020kjk,Lamm:2019bik,Brower:2020huh,Kreshchuk:2020dla}. In this work, we will focus on step (3), assuming that steps (1) and (2) have been achieved. It should be remarked that, in step (3), the three objects to be calculated could be ultraviolet divergent, meaning that
their individual values blow up in the continuum limit. However, the scattering amplitude, as a physical observable, remains a finite constant when the continuum limit is taken.
The large cancellation in the continuum limit among the components in the LSZ reduction formula could potentially cause problems on numerical stability in practical calculations.   
We leave the detailed study of the approach to the continuum limit of the LSZ reduction formula for the future.

\section{The quantum algorithm} 
Here we propose a quantum algorithm to compute the three objects involved in the LSZ reduction formula Eq.~(\ref{eq:LSZ}), namely the connected
 $n$-point function, the propagator, and the field normalization. Accordingly to Eq.~(\ref{eq:R}), the field normalization $R$ involves the field operator $\phi$ evaluated at $x=0$ sandwiched between the vacuum and a single-particle state with zero spatial momentum ($\bm{p}=0$). Since no time evolution of the field operator is involved, the value of $R$ can be readily determined
once the vacuum and the  single-particle  state are obtained. To obtain the vacuum and the  single-particle  state, one can employ the quantum algorithm proposed in Ref.~\cite{Li:2021kcs}, which shows that both the vacuum and the single-particle  state can be obtained efficiently with the quantum alternating operator ansatz (QAOA) and the quantum-number-resolving variational quantum eigensolver~\cite{farhi2014quantum,QAOA,wiersema-PRXQuantum2020,nakanishi_19}. It should be noted that, only states with
zero spatial momentum are involved in our formalism.\footnote{In this work, we only consider massive particles, for which a rest frame exists.} Since these states are translation invariant, the QAOA can be applied easily: one simply uses input reference states and alternating operators that are constructed to be translation invariant. Next, we need to compute the connected $n$-point function and the propagator, for which again we can use the quantum
algorithm proposed in Ref.~\cite{Li:2021kcs} developed for the evaluation of parton distribution functions (PDFs), based on the general method
introduced in Ref.~\cite{Pedernales:2014}. In Refs.~\cite{Li:2021kcs} and~\cite{Li:2022lyt}, with simulations on classical hardware, it is shown that with such a quantum algorithm the PDF and the light-cone distribution amplitude of the one-flavor Gross-Neveu model can be
obtained with good accuracy with only 18 and 14 qubits, respectively. 

Our approach to the quantum computation of scattering amplitudes differs in many ways from the JLP formalism. 
The essential difference is that, the JLP formalism is a direct Hamiltonian simulation of the scattering process, for which the 
outgoing particle states are unknown; while in our approach we specify the ingoing and outgoing states, and aim at calculating the amplitude of the 
specified scattering process by evaluating relevant correlation functions. In this regard, computational cost is reduced in our approach in two
ways. First, in our framework, the contraint Eq.~(\ref{eq:constraint_2}) is relaxed and a smaller lattice is allowed. Second, since there is no adiabatic
turn on or turn off of the coupling constant in our formalism, there is no associated extra time evolution and corrections to broadening of wave packets, and thus the associated theoretical errors can be avoided. 

It should be noted that, in the JLP formalism, the outgoing state is unknown, and measurements of momentum-space occupation numbers are performed on the final state in order to extract
information on the outgoing particles. The scattering amplitude of a specific scattering process is obtained only after enough statistics
is obtained for the specific process. In our approach, the outgoing state is known.
However, this does not necessarily mean that our approach involves
a fewer number of gates. In fact, according to Eq.~(\ref{eq:G_m_n}), we have to evaluate the position-space connected $n$-point function
at every spacetime point and then perform a Fourier transform. We can estimate the computational complexity  
in our approach as follows. 
Suppose we have $N$ lattice sites and $T$ temporal sites, with $d$ spatial dimensions. Suppose we need $n_{q}$ qubits at each lattice site, then the total number of qubits for storage of the state is $ n_{q}N$. To prepare the vacuum or a one-particle state with the QAOA and the variational 
quantum eigensolver, the number of gate operations is $\mathcal{O}(n_{q}N )$~\cite{wiersema-PRXQuantum2020,Li:2021kcs}. The complexity of preparing $n$ one-particle states is therefore $\mathcal{O}(n n_{q} N )$. With the Trotter formula, conservatively the time evolution is estimated to cost $\mathcal{O}(n_{q} N^2 T)$ operations~\cite{nielsen_chuang_2010,Byrnes:2005qx}.  
From this, we infer that each evaluation of the position-space $n$-point function has complexity $\mathcal{O}(nn_{q}N^2T)$~\cite{Pedernales:2014}, and so the complexity for evaluating 
the position-space $n$-point function at all spacetime points is $\mathcal{O}(nn_{q}N^2T(NT)^{n-1})=\mathcal{O}(nn_{q}N^{n+1}T^{n})$.
Assuming $\mathcal{O}(T)=\mathcal{O}(N^{1/d})$, the subsequent Fourier transform can be done efficiently using the multidimensional quantum Fourier transform with complexity $\mathcal{O}(\frac{1}{(n-1)(d+1)}\log^2\left(N^{(n-1)(d+1)/d}\right))$~\cite{Pfeffer:2023yhb}, which is negligible compared to $\mathcal{O}(nn_{q}N^{n+1}T^{n})$.
To get the connected part of the $n$-point function, we also have to calculate its disconnected part, the evaluation of which according to what we have just derived has complexity
$\mathcal{O}\left(\sum_{k=1}^{n-1}\left(\begin{array}{c}n\\k\end{array}\right)kn_qN^{k+2}T^{k+1}\right)\le\mathcal{O}\left(2^nnn_qN^{n+1}T^{n}\right)$. Similarly, evaluating the $n$ propagators in Eq.~(\ref{eq:LSZ})
costs $\mathcal{O}(nn_{q}N^{3}T^{2})$ operations.
Consequently, the overall complexity in our approach is $\mathcal{O}(2^nnn_{q}N^{n+1}T^{n})$,
 which is exponential in $n$.
In the JLP formalism, it was shown that the complexity scales with $n$ polynomially~\cite{Jordan:2012xnu}. Therefore, our approach
is ideal only when the number of external particles is small, e.g. in $2\to 2$ scatterings.

With the estimate of complexity above, we can see how the complexity scales with the largest energy scale in the scattering process, denoted by $\Lambda_{\rm max}$.
In general, in order to satisfy constraint Eq.~(\ref{eq:constraint_1}), both $N$ and $T$ are taken to be proportional to $\Lambda_{\rm max}$. For fermionic degrees of freedom,
$n_q$ is fixed and does not depend on $\Lambda_{\rm max}$. 
For bosonic degrees of freedom, $n_q$ is chosen as a finite number which is large enough so that modes of energy or momentum
of order $\Lambda_{\rm max}$ are properly represented. It is found that $n_q$ can be taken as $\mathcal{O}(\log \Lambda_{\rm max})$ for scalar fields~\cite{Klco:2018zqz} and gauge fields~\cite{Byrnes:2005qx},
if we estimate the maximum value of the field strength tensor as $\sim \Lambda_{\rm max}^2$.
Therefore, the complexity scales at most as $\sim\Lambda_{\rm max}^{2n+1}\log\Lambda_{\rm max}$, which is polynomial in $\Lambda_{\rm max}$. A polynomial dependence of the complexity 
on energy was also obtained in the JLP formalism~\cite{Jordan:2012xnu,Jordan:2014tma}.

An important feature of our approach using the LSZ reduction formula is that bound states are allowed as incoming or outgoing particles. This is because the interaction coupling constant is never turned off in our approach, as opposed to the JLP formalism.   
In Eqs.~(\ref{eq:G_m_n})--(\ref{eq:R}), the field operator $\phi$ is not necessarily a fundamental field of the theory. In fact, any
operator which has the same quantum numbers as the external particle $h$ can be used. 
For instance, in a theory with only a spin-$1/2$ fundamental field $\psi$, there might exist a spin-$0$ scalar bound state $h$ made of a fundamental fermion and its antiparticle. One can then simply take  the composite  operator $\phi=\bar{\psi}\psi$ as the operator which annihilates $h$ in the LSZ reduction formula for scattering processes involving $h$ as external particles. 
This is an ideal feature of our formalism, since in the most interesting potential application of quantum computing in particle physics, namely quantum chromodynamics, all incoming and outgoing particles are bound states owing to quark confinement. Our framework is therefore most useful for
scattering processes involving a small number of composite particles in a strongly coupled theory, such as $2\to 2$ scatterings of pions or nucleons in quantum chromodynamics, for which
first-principle calculations are currently only possible below the three-hadron thresholds~\cite{Luscher:1990ux,Briceno:2017max,Andersen:2018mau}.

\section{Polology in the one-flavor Gross-Neveu model}
With simulations using the proposed quantum algorithm on classical hardware, we can demonstrate the feature of poles of the propagator and the connected four-point function in a simple model, the (1+1)-dimensional one-flavor Nambu-Jona-Lasinio model~\cite{Nambu:1961tp,Nambu:1961fr}, also known as the one-flavor Gross-Neveu model~\cite{Gross:1974jv}. The Lagrangian of this model is given by
\begin{equation}\label{eq:LCDA}
	\mathcal{L}=\bar{\psi} (i\gamma^\mu \partial_\mu-m_q)\psi +g(\bar{\psi} \psi)^2\,,
\end{equation}
where $\psi$ is a Dirac field in 1+1 dimensions, which we will refer to as the quark field, and $m_q$ and $g$ are the bare quark mass and the bare coupling constant, respectively. Both $m_q$ and $g$ are free
parameters. We choose the values of them in such a way that the particle states $h$ we consider below have masses $m_h$ satisfying $\frac{\pi}{L}<m_h<\frac{\pi}{a}$, where $a$ and $L$ are the lattice spacing and the lattice size, respectively. In the following, we take 
$m_qa=0.84$ and $g=0.40$.

Consider the propagator of the quark field,
\begin{align}
K_\psi(p)
&=\int d^2x\,e^{ip\cdot x} \langle \Omega|T\{\psi(x) \bar{\psi}(0)\}|\Omega\rangle\,.\label{eq:K_NJL}
\end{align}
Similar to Eq.~(\ref{eq:2_pole}), near the mass shell of a particle state $h$ which has the same quantum numbers as the quark field, we have
\begin{align}
K_\psi(p)\overset{p^2\to m_h^2}{\longrightarrow}\frac{iR_h\,(\slashed{p}+m_h)}{p^2-m_h^2+i\epsilon}\,,\label{eq:2_pole_NJL}
\end{align}
where $R_h>0$ is the field normalization defined by
\begin{align}
\langle \Omega|\psi(0)|h(p)\rangle =\sqrt{R_h}\,u(p)\,,\label{eq:R_psi}
\end{align}
where $u(p)$ is the positive-energy solution to the free Dirac equation $(\slashed{p}-m_h)u(p)=0$, normalized to $\bar{u}(p)u(p)=2m_h$.
The field normalization $R_h$ can be calculated using our proposed method by taking $\bm{p}=0$ in Eq.~(\ref{eq:R_psi}).

\begin{figure}[htbp]
	\centering
	\includegraphics[width=0.5 \textwidth]{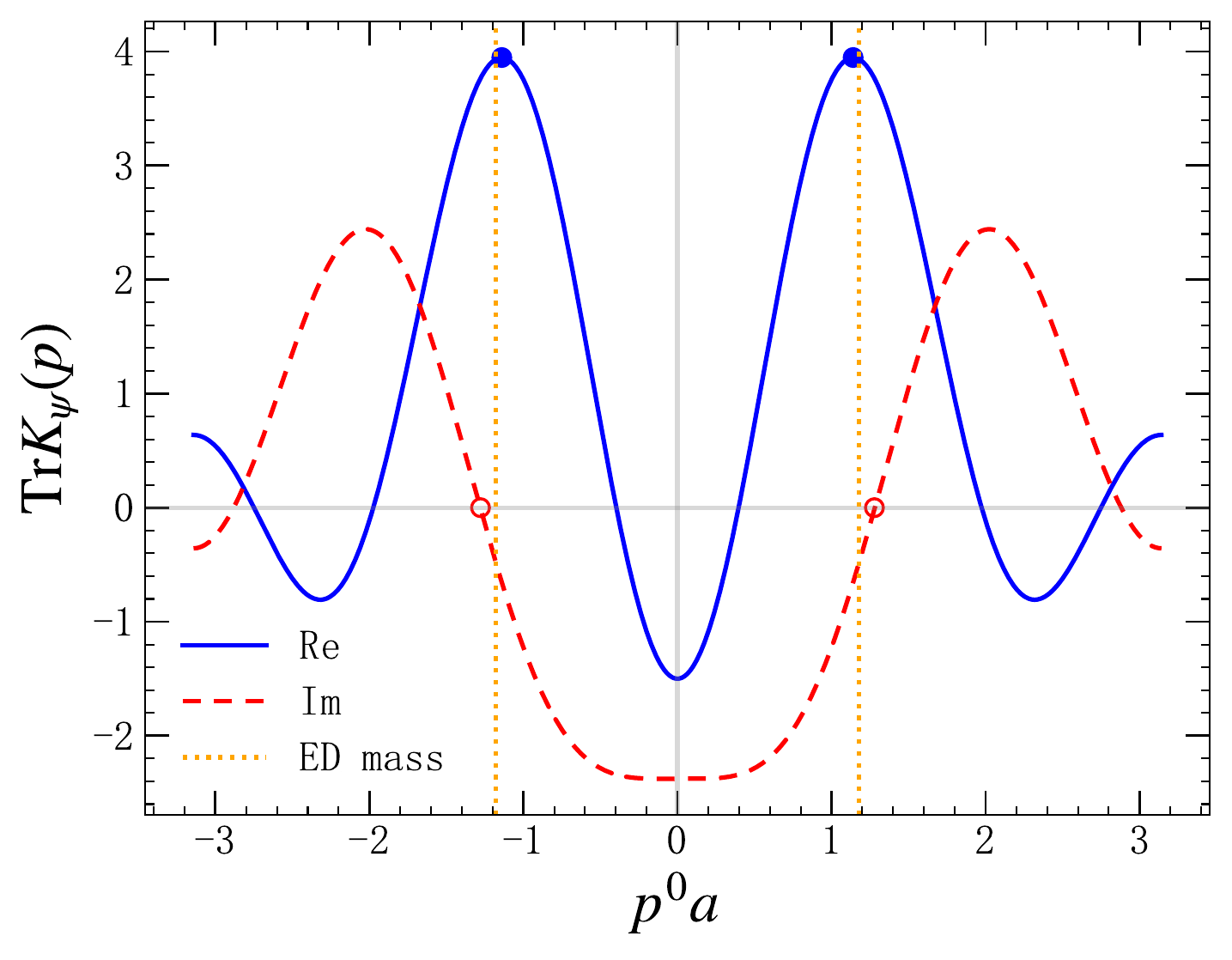}
	\caption{Real part (solid line) and imaginary part (dashed line) of ${\rm Tr}K_\psi(p)$ in the one-flavor Gross-Neveu model as a function of $p^0a$ with $p^1=0$, simulated with the proposed quantum algorithm on classical hardware. The solid blobs on the solid line and the hollow circles
on the dashed line, respectively, show the peaks of the real part and the zeros of the imaginary part of ${\rm Tr}K_\psi(p)$ corresponding to the
pole structure due to the quark. The dotted vertical lines show the locations $p^0a=\pm m_h a$ with the quark mass $m_h$ obtained from exact 
diagonalization of the discretized Hamiltonian.}
\label{fig:TrK}
\end{figure}

According to Eq.~(\ref{eq:2_pole_NJL}), when we take $p^1=0$ and treat $K_\psi(p)$ as a function of $p^0$, $K_\psi(p)$ should have poles at $p^0=\pm m_h$. We evaluate $K_\psi(p)$ as a function $p^0$, with $p^1$ taken to be zero, with the proposed quantum algorithm using classical hardware.
The calculation is performed on a desktop workstation with $16$ cores, using opensource packages \textsc{QuSpin}~\cite{quspin} and 
\textsc{projectQ}~\cite{Steiger2018projectqopensource}, with $14$ qubits (seven lattice sites),
using seven temporal sites, with the temporal spacing taken to be the same as the lattice spacing. We follow
the mapping of the Gross-Neveu model onto qubits  and  the method to evaluate the correlation function with a quantum algorithm as discussed in Ref.~\cite{Li:2021kcs}.
To obtain $K_\psi(p)$, we first calculate the integrand in Eq.~(\ref{eq:K_NJL}) in position space and then implement a discrete Fourier transform to perform the Fourier integral. 
In order to show the main features of our result, we present our result for ${\rm Tr}K_\psi(p)$, which according to
Eq.~(\ref{eq:2_pole_NJL}) takes the following form near a particle mass shell:
\begin{align}
{\rm Tr}K_\psi(p)\overset{p^2\to m_h^2}{\longrightarrow}
\frac{2R_hm_h\left[\epsilon+i(p^2-m_h^2)\right]}{(p^2-m_h^2)^2+\epsilon^2}\,.\label{eq:Tr_pole_NJL}
\end{align}
Equation~(\ref{eq:Tr_pole_NJL}) shows that ${\rm Re}\left[{\rm Tr}K_\psi(p)\right]\to +\infty$ and 
${\rm Im}\left[{\rm Tr}K_\psi(p)\right]\to 0$ as $p^2\to m_h^2$.
Figure~\ref{fig:TrK}
shows our results for the real part (solid line) and the imaginary part (dashed line) of ${\rm Tr}K_\psi(p)$ as a function of $p^0a$. The peaks of the real part at $p^0a=\pm 1.14$ (solid blobs) correspond to the poles from the lowest-lying state with the same quantum numbers as the quark field, as is verified by solving for the mass spectrum with direct numerical diagonalization of the discretized Hamiltonian, which gives $m_ha=1.18$ 
($\pm m_ha$ shown by dotted vertical lines). 
This state can be interpreted as a quark.\footnote{Note that in this model there is no quark confinement.} In the continuum limit, a pole corresponds to a peak of infinite height, while in the discretized model we consider here the peaks have finite height.

According to Eq.~(\ref{eq:Tr_pole_NJL}), when viewed as a function of $p^0$ with $p^1=0$, ${\rm Im}\left[{\rm Tr}K_\psi(p)\right]$
has zeros at $p^0=m_h$ and $p^0=-m_h$, at which points the derivative of ${\rm Im}\left[{\rm Tr}K_\psi(p)\right]$ with respect to $p^0$ is positive
and negative, respectively. These features are exhibited in our result for the imaginary part of ${\rm Tr}K_\psi(p)$, shown by the dashed line in
Fig.~\ref{fig:TrK}. In Fig.~\ref{fig:TrK}, the dashed line has positive and negative derivative at the zeros at $p^0a= 1.28$ and $p^0a=-1.28$ (hollow circles), respectively. These zeros are close to $\pm m_h a$, and are expected to be there by virtue of Eq.~(\ref{eq:Tr_pole_NJL}) in the continuum theory.

\begin{figure}[htbp]
	\centering
	\includegraphics[width=0.5 \textwidth]{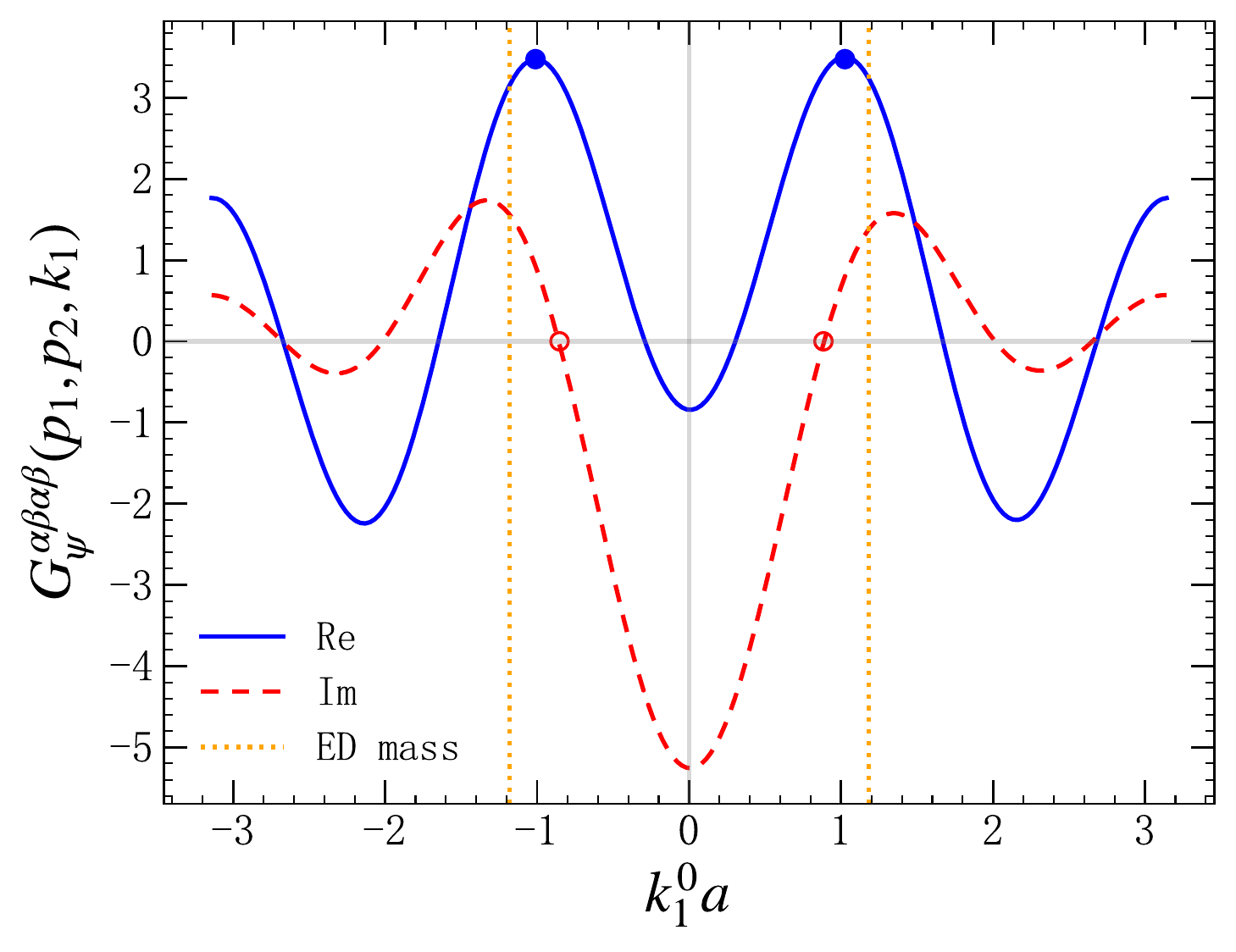}
	\caption{Real part (solid line) and imaginary part (dashed line) of $G^{\alpha\beta\alpha\beta}_\psi(p_1,p_2,k_1)$ in the one-flavor Gross-Neveu model as a function of $k_1^0a$, with $k_1=(k_1^0,0),\,p_1=(0,0),\,p_2=(k_1^0,\pi/a)$, simulated with the proposed quantum algorithm on classical hardware. The solid blobs on the solid line and the hollow circles
on the dashed line, respectively, show the peaks of the real part and the zeros of the imaginary part of $G^{\alpha\beta\alpha\beta}_\psi(p_1,p_2,k_1)$ corresponding to the pole structure due to the quark. The dotted vertical lines show the locations $p^0a=\pm m_h a$ with the quark mass $m_h$ obtained from exact diagonalization of the discretized Hamiltonian. }
\label{fig:TrG}
\end{figure}

For the $2\to 2$ scattering of a quark and an antiquark, ${q}(\bm{k_1})\bar{q}(\bm{k_2})\to {q}(\bm{p_1})\bar{q}(\bm{p_2})$, one has to calculate the following connected four-point function: 
\begin{align}
&G^{\alpha\beta\gamma\delta}_\psi(p_1,p_2,k_1) \nonumber\\
&=\,\int d^2x_1 d^2x_2 d^2y_1\, e^{i(p_1\cdot x_1+p_2\cdot x_2-k_1\cdot y_1)}\nonumber\\
&\quad\,\times\langle\Omega|\psi^\alpha(x_1)\bar{\psi}^\beta(x_2)\bar{\psi}^\gamma(y_1){\psi}^\delta(0)|\Omega\rangle_{\rm con}\,.
\end{align} 
Similar to $K_\psi(p)$, the connected four-point function $G^{\alpha\beta\gamma\delta}_\psi(p_1,p_2,k_1)$ is expected to have a pole structure of $\frac{1}{k_1^2-m_h^2+i\epsilon}$ when $k_1$ is close to the mass shell of particle $h$ which has the same quantum numbers as the quark field. To demonstrate this feature, with the same method as we evaluate $K_\psi(p)$, we evaluate $G^{\alpha\beta\gamma\delta}_\psi(p_1,p_2,k_1)$ as a function of $k_1^0$ with the external-leg momenta set to $k_1=(k_1^0,0),\,p_1=(0,0),\,p_2=(k_1^0,\pi/a)$. This setting of external-leg momenta makes sure that
$k_2,p_1,p_2$ are off shell. 
Figure~\ref{fig:TrG}
shows our results for the real part (solid line) and the imaginary part (dashed line) of $G^{\alpha\beta\alpha\beta}_\psi(p_1,p_2,k_1)$  as a function of $k_1^0a$. Similar to the case of the propagator, the peaks of the real part of $G^{\alpha\beta\alpha\beta}_\psi(p_1,p_2,k_1)$ at $k_1^0a=\pm 1.01$ (solid blobs) correspond to the poles from the lowest-lying state with the same quantum numbers as the quark field, namely the quark. The imaginary part of $G^{\alpha\beta\alpha\beta}_\psi(p_1,p_2,k_1)$, shown by the dashed line in Fig.~\ref{fig:TrG}, has positive and negative 
derivative, respectively, at the zeros at $k_1^0a= 0.87$ and $k_1^0a=-0.87$ (hollow circles). These zeros, being close to $\pm m_h a$, are expected by the pole structure of the connected four-point function in the continuum theory.

\begin{figure}[htbp]
	\centering
	\includegraphics[width=0.5 \textwidth]{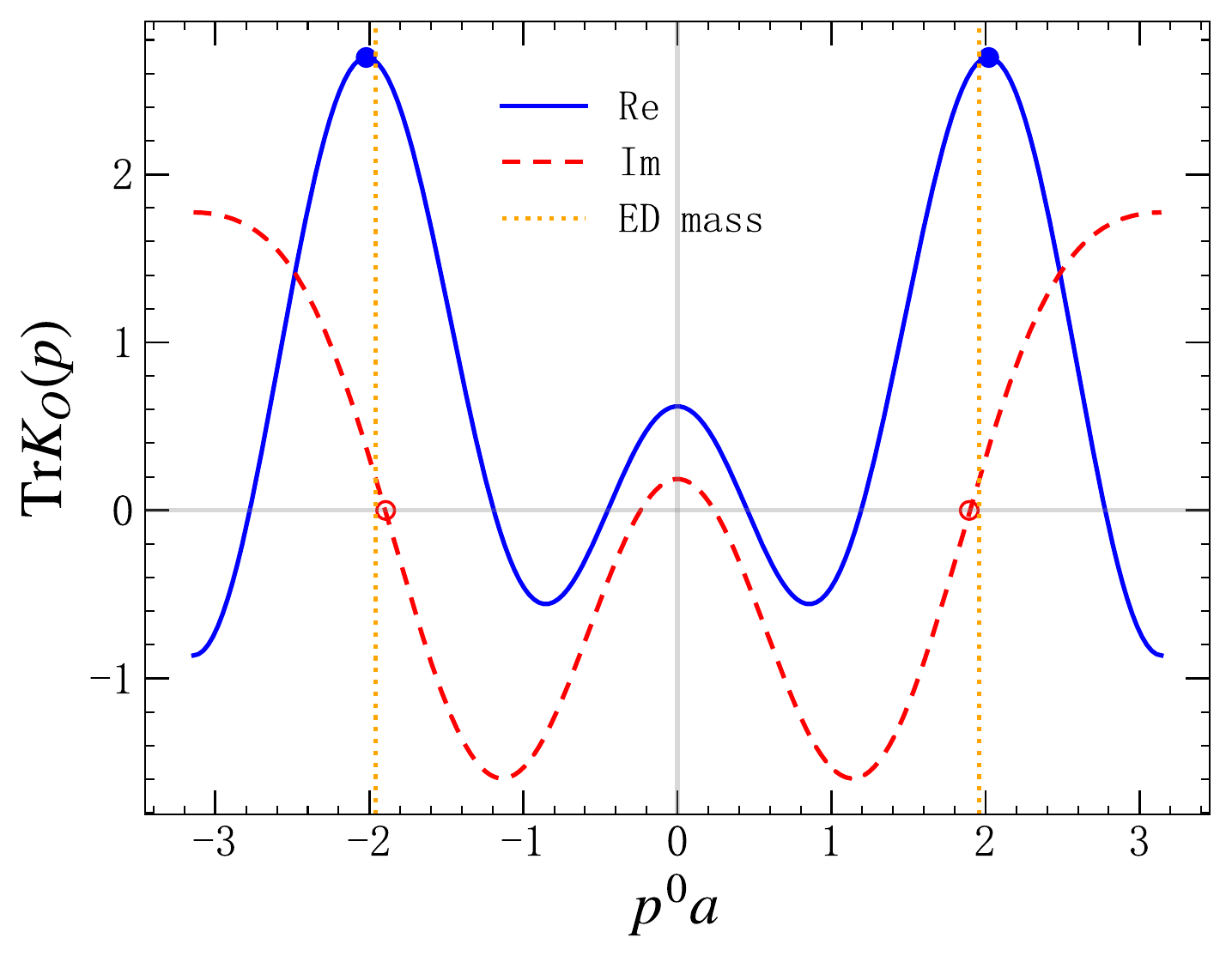}
	\caption{Real part (solid line) and imaginary part (dashed line) of ${\rm Tr}K_O(p)$ in the one-flavor Gross-Neveu model as a function of $p^0a$ with $p^1=0$, simulated with the proposed quantum algorithm on classical hardware. The solid blobs on the solid line and the hollow circles on the dashed line, respectively, show the peaks of the real part and the zeros of the imaginary part of ${\rm Tr}K_O(p)$ corresponding to the
pole structure due to the quark-antiquark bound state. The dotted vertical lines show the locations $p^0a=\pm m_{h_O} a$ with the bound-state mass $m_{h_O}$ obtained from exact 
diagonalization of the discretized Hamiltonian.}
	\label{fig:TrO}
\end{figure}

In order to demonstrate the power of the LSZ reduction formula in handling scatterings of bound-state particles, we also simulate the propagator of the composite operator $O(x)=\bar{\psi}(x)\psi(x)$, given by
\begin{align}
K_O(p)
&=\int d^2x\,e^{ip\cdot x} \langle \Omega|T\{O(x) O(0)\}|\Omega\rangle_{\rm con}\,.\label{eq:K_O}
\end{align}
Figure~\ref{fig:TrO} shows our results for the real part (solid line) and the imaginary part (dashed line) of ${\rm Tr}K_O(p)$ as a function of $p^0a$, with $p^1=0$. The peaks of the real part at $p^0a=\pm 2.02$ (solid blobs) correspond to the poles from the second lowest-lying state $h_O$ with the same quantum numbers as the vacuum, as is verified by solving for the mass spectrum with direct numerical diagonalization, which gives 
$m_{h_O}a=1.96$ ($\pm m_{h_O}a$ shown by dotted vertical lines). This state can be interpreted as a quark-antiquark bound state.
As expected again from the pole structure of $K_O(p)$ in the continuum theory, the imaginary part has positive and negative derivative, 
respectively, at the zeros at $p^0a= 1.89$ and $p^0a=-1.89$ (hollow circles), which are close to $\pm m_{h_O}a$.
Note that the peak of the real part at $p^0a=0$ does not correspond to any single-particle pole, since the pole structure due to such a massless state would imply that the imaginary part has a zero at $p^0a=0$ which is a local minimum, in contrary to the local maximum we found.

This simple example shows that our proposed quantum algorithm succeeds in recovering the expected pole structure of both the propagator and the connected $n$-point function, which is crucial to the implementation of the LSZ reduction formula. 

It should be noted that, although we find that both the quark propagator and the connected four-point function exhibit convergent behavior when the number of qubits is increased to $14$ from a smaller number, the speeds of their convergence are quite different, with the propagator converging faster. For $2\to 2$ quark-antiquark scattering, since the scattering amplitude involves four powers of the ratio of peak values of the connected 
four-point function and the quark propagator, its convergence requires that both the quark propagator and the connected four-point function reach a similar speed of convergence. In our initial attempt to calculate this scattering amplitude, owing to the limited number of qubits one can use in simulations on classical hardware with reasonable computational time cost, a satisfactory convergence has not been observed. However, with the quantum advantage, in a quantum computer with more than a hundred qubits in near future, we believe that a convergent result for the scattering amplitude can be expected for a (1+1)-dimensional model.  

\section{Conclusions}
In this work, we proposed a new framework for evaluating scattering amplitudes in quantum field theory on quantum computers in a fully nonperturbative way. The framework was based on the LSZ reduction formula, which relates scattering amplitudes to correlation functions. In this framework, as opposed to a direct Hamiltonian simulation of the scattering process, no preparation of wave packets of incoming particles is required, and one only has to prepare one-particle states of zero momentum. The framework is capable of incorporating scatterings of bound-state particles, and is
ideal for scatterings that involve a small number of particles. This framework is expected to have potential applications in exclusive processes in a
strongly coupled theory, such as $2\to 2$ scatterings of pions or nucleons. 
As a proof of concept, in a simple model, the one-flavor Gross-Neveu model, we demonstrated by simulations on classical hardware that the propagator and the connected four-point function obtained
from the quantum algorithm has the desired pole structure crucial to the implementation of the LSZ reduction formula.

\begin{acknowledgments}
This work is supported by the Guangdong Major Project of Basic and Applied Basic Research No. 2020B0301030008 and No. 2022A1515010683, and by the National Natural Science Foundation of China (NSFC) under Grants No. 12035007 and No. 12022512.
\end{acknowledgments}


\bibliographystyle{h-physrev5}

\begin{thebibliography}{10}

\bibitem{Feynman:1949zx}
R.~P. Feynman,
Space-Time Approach to Quantum Electrodynamics,
\newblock Phys. Rev. {\bf 76}, 769 (1949).

\bibitem{Eden:1966dnq}
R.~J. Eden, P.~V. Landshoff, D.~I. Olive, and J.~C. Polkinghorne,
\newblock {\em {The Analytic S-Matrix}} (Cambridge University Press, Cambridge, England, 1966).

\bibitem{Bern:1994fz}
Z.~Bern, L.~J. Dixon, and D.~A. Kosower,
One-loop corrections to two-quark three-gluon amplitudes,
\newblock Nucl. Phys. B {\bf 437}, 259 (1995), arXiv:hep-ph/9409393.

\bibitem{Britto:2005fq}
R.~Britto, F.~Cachazo, B.~Feng, and E.~Witten,
Direct Proof of the Tree-Level Scattering Amplitude Recursion Relation in Yang-Mills Theory,
\newblock Phys. Rev. Lett. {\bf 94}, 181602 (2005), arXiv:hep-th/0501052.

\bibitem{Bern:2007dw}
Z.~Bern, L.~J. Dixon, and D.~A. Kosower,
On-shell methods in perturbative QCD,
\newblock Ann. Phys. (Amsterdam) {\bf 322}, 1587 (2007), arXiv:0704.2798.

\bibitem{ALEPH:2005ab}
S.~Schael {\em et~al.}, (ALEPH, DELPHI, L3, OPAL, SLD, LEP Electroweak Working Group, SLD Electroweak
  Group, SLD Heavy Flavour Group Collaborations),
Precision electroweak measurements on the Z resonance,
\newblock Phys. Rep. {\bf 427}, 257 (2006), arXiv:hep-ex/0509008.

\bibitem{Anastasiou:2016cez}
C.~Anastasiou {\em et~al.},
High precision determination of the gluon fusion Higgs boson cross-section at the LHC,
\newblock J. High Energy Phys. {\bf 05}, 058 (2016), arXiv:1602.00695.

\bibitem{ATLAS:2019zci}
G.~Aad {\em et~al.}, (ATLAS Collaboration)
Measurement of the transverse momentum distribution of Drell-Yan lepton pairs in proton-proton collisions at $\sqrt{s}=13$ TeV with the ATLAS detector,
\newblock Eur. Phys. J. C {\bf 80}, 616 (2020), arXiv:1912.02844.

\bibitem{Alexandru:2016gsd}
A.~Alexandru, G.~Basar, P.~F. Bedaque, S.~Vartak, and N.~C. Warrington,
Monte Carlo Study of Real Time Dynamics on the Lattice,
\newblock Phys. Rev. Lett. {\bf 117}, 081602 (2016), arXiv:1605.08040.

\bibitem{Jordan:2012xnu}
S.~P. Jordan, K.~S.~M. Lee, and J.~Preskill,
Quantum Algorithms for Quantum Field Theories,
\newblock Science {\bf 336}, 1130 (2012), arXiv:1111.3633.

\bibitem{Jordan:2011ci}
S.~P. Jordan, K.~S.~M. Lee, and J.~Preskill,
Quantum computation of scattering in scalar quantum field theories,
\newblock Quant. Inf. Comput. {\bf 14}, 1014 (2014), arXiv:1112.4833.

\bibitem{Jordan:2014tma}
S.~P. Jordan, K.~S.~M. Lee, and J.~Preskill,
Quantum Algorithms for Fermionic Quantum Field Theories,
\newblock arXiv:1404.7115.

\bibitem{Bauer:2022hpo}
C.~W. Bauer {\em et~al.},
Quantum Simulation for High-Energy Physics,
\newblock PRX Quantum {\bf 4}, 027001 (2023), arXiv:2204.03381.

\bibitem{Martinez:2016yna}
E.~A. Martinez {\em et~al.},
Real-time dynamics of lattice gauge theories with a few-qubit quantum computer,
\newblock Nature (London) {\bf 534}, 516 (2016), arXiv:1605.04570.

\bibitem{Hu:2019hrf}
Z.~Hu, R.~Xia, and S.~Kais,
A quantum algorithm for evolving open quantum dynamics on quantum computing devices,
\newblock Sci. Rep. {\bf 10}, 3301 (2020), arXiv:1904.00910.

\bibitem{Bauer:2019qxa}
B.~Nachman, D.~Provasoli, W.~A. de~Jong, and C.~W. Bauer,
Quantum Algorithm for High Energy Physics Simulations,
\newblock Phys. Rev. Lett. {\bf 126}, 062001 (2021), arXiv:1904.03196.

\bibitem{DeJong:2020riy}
W.~A. de~Jong, M.~Metcalf, J.~Mulligan, M.~Ploskon, F.~Ringer, and X.~Yao,
Quantum simulation of open quantum systems in heavy-ion collisions,
\newblock Phys. Rev. D {\bf 104}, L051501 (2021), arXiv:2010.03571.

\bibitem{Zhou:2021kdl}
Z.-Y. Zhou {\em et~al.},
Thermalization dynamics of a gauge theory on a quantum simulator,
\newblock Science {\bf 377}, 311 (2022), arXiv:2107.13563.

\bibitem{deJong:2021wsd}
W.~A. de~Jong, and K.~Lee, J.~Mulligan, M.~Ploskon, F.~Ringer, and X.~Yao,
Quantum simulation of nonequilibrium dynamics and thermalization in the Schwinger model,
\newblock Phys. Rev. D {\bf 106}, 054508 (2022), arXiv:2106.08394.

\bibitem{Bepari:2021kwv}
K.~Bepari, S.~Malik, M.~Spannowsky, and S.~Williams,
Quantum walk approach to simulating parton showers,
\newblock Phys. Rev. D {\bf 106}, 056002 (2022), arXiv:2109.13975.

\bibitem{Atas:2022dqm}
Y.~Y. Atas {\em et~al.},
Simulating one-dimensional quantum chromodynamics on a quantum computer: Real-time evolutions of tetra- and pentaquarks,
\newblock Phys. Rev. Res. {\bf 5} 033184 (2023), arXiv:2207.03473.

\bibitem{Yao:2022eqm}
X.~Yao,
Quantum Simulation of Light-Front QCD for Jet Quenching in Nuclear Environments,
\newblock (2022), arXiv:2205.07902.

\bibitem{Lu:2018pjk}
H.-H. Lu {\em et~al.},
Simulations of subatomic many-body physics on a quantum frequency processor,
\newblock Phys. Rev. A {\bf 100}, 012320 (2019), arXiv:1810.03959.

\bibitem{Lamm:2019uyc}
H.~Lamm, S.~Lawrence, and Y.~Yamauchi (NuQS Collaboration),
Parton physics on a quantum computer,
\newblock Phys. Rev. Res. {\bf 2}, 013272 (2020), arXiv:1908.10439.

\bibitem{Mueller:2019qqj}
N.~Mueller, A.~Tarasov, and R.~Venugopalan,
Deeply inelastic scattering structure functions on a hybrid quantum computer,
\newblock Phys. Rev. D {\bf 102}, 016007 (2020), arXiv:1908.07051.

\bibitem{Roggero:2019myu}
A.~Roggero, A.~C.~Y. Li, J.~Carlson, R.~Gupta, and G.~N. Perdue,
Quantum computing for neutrino-nucleus scattering,
\newblock Phys. Rev. D {\bf 101}, 074038 (2020), arXiv:1911.06368.

\bibitem{Echevarria:2020wct}
M.~G. Echevarria, I.~L. Egusquiza, E.~Rico, and G.~Schnell,
Quantum simulation of light-front parton correlators,
\newblock Phys. Rev. D {\bf 104}, 014512 (2021), arXiv:2011.01275.

\bibitem{Kreshchuk:2020kcz}
M.~Kreshchuk {\em et~al.},
Light-Front Field Theory on Current Quantum Computers,
\newblock Entropy {\bf 23}, 597 (2021), arXiv:2009.07885.

\bibitem{Bauer:2021gup}
C.~W. Bauer, B.~Nachman, and M.~Freytsis, 
Simulating Collider Physics on Quantum Computers Using Effective Field Theories,
\newblock Phys. Rev. Lett. {\bf 127}, 212001 (2021), arXiv:2102.05044.

\bibitem{Atas:2021ext}
Y.~Y. Atas {\em et~al.},
$SU(2)$ hadrons on a quantum computer via a variational approach,
\newblock Nat. Commun. {\bf 12}, 6499 (2021), arXiv:2102.08920.

\bibitem{Li:2021kcs}
T.~Li, X.~Guo, W.~K.~Lai, X.~Liu, E.~Wang, H.~Xing, D.~B.~Zhang, and S.~L.~Zhu (QuNu Collaboration),
Partonic collinear structure by quantum computing,
\newblock Phys. Rev. D {\bf 105}, L111502 (2022), arXiv:2106.03865.

\bibitem{Li:2022lyt}
T.~Li, X.~Guo, W.~K.~Lai, X.~Liu, E.~Wang, H.~Xing, D.~B.~Zhang, and S.~L.~Zhu (QuNu Collaboration),
Exploring light-cone distribution amplitudes from quantum computing,
\newblock Sci. China Phys. Mech. Astron. {\bf 66}, 281011 (2023), arXiv:2207.13258.

\bibitem{Gallimore:2022hai}
D.~Gallimore and J.~Liao,
Quantum computing for heavy quarkonium spectroscopy,
\newblock Phys. Rev. D {\bf 107}, 074012 (2023), arXiv:2202.03333.

\bibitem{Czajka:2021yll}
A.~M. Czajka, Z.-B. Kang, H.~Ma, and F.~Zhao,
Quantum simulation of chiral phase transitions,
\newblock J. High Energy Phys. {\bf 08}, 209 (2022), arXiv:2112.03944.

\bibitem{Czajka:2022plx}
A.~M. Czajka, Z.-B. Kang, Y.~Tee, and F.~Zhao,
Studying chirality imbalance with quantum algorithms,
\newblock arXiv:2210.03062.

\bibitem{Xie:2022jgj}
X.~D.~Xie, X.~Guo, H.~Xing, Z.~Y.~Xue, D.~B.~Zhang, and S.~L.~Zhu (QuNu Collaboration),
Variational thermal quantum simulation of the lattice Schwinger model,
\newblock Phys. Rev. D {\bf 106}, 054509 (2022), arXiv:2205.12767.

\bibitem{LSZ}
H.~Lehmann, H.~Symanzik, and W.~Zimmerman,
Zur Formulierung quantisierter Feldtheorien,
\newblock Nuovo Cimento {\bf 1}, 205 (1955).

\bibitem{Briceno:2020rar}
R.~A. Brice\~no, J.~V. Guerrero, M.~T. Hansen, and A.~M. Sturzu,
Role of boundary conditions in quantum computations of scattering observables,
\newblock Phys. Rev. D {\bf 103}, 014506 (2021), arXiv:2007.01155.

\bibitem{Kogut:1974ag}
J.~B. Kogut and L.~Susskind,
Hamiltonian formulation of Wilson's lattice gauge theories,
\newblock Phys. Rev. D {\bf 11}, 395 (1975).

\bibitem{Kogut:1982ds}
J.~B. Kogut,
The lattice gauge theory approach to quantum chromodynamics,
\newblock Rev. Mod. Phys. {\bf 55}, 775 (1983).

\bibitem{Anishetty:2009nh}
R.~Anishetty, M.~Mathur, and I.~Raychowdhury,
Prepotential formulation of $SU(3)$ lattice gauge theory,
\newblock J. Phys. A {\bf 43}, 035403 (2010), arXiv:0909.2394.

\bibitem{Raychowdhury:2019iki}
I.~Raychowdhury and J.~R. Stryker,
Loop, string, and hadron dynamics in $SU(2)$ Hamiltonian lattice gauge theories,
\newblock Phys. Rev. D {\bf 101}, 114502 (2020), arXiv:1912.06133.

\bibitem{Buser:2020cvn}
A.~J. Buser, H.~Gharibyan, M.~Hanada, M.~Honda, and J.~Liu,
Quantum simulation of gauge theory via orbifold lattice,
\newblock J. High Energy Phys. {\bf 09}, 034 (2021), arXiv:2011.06576.

\bibitem{Jordan_Wigner}
P.~Jordan and E.~P. Wigner,
{\"U}ber das Paulische {\"A}quivalenzverbot,
\newblock Z. Phys. {\bf 47}, 631 (1928).

\bibitem{Bravyi}
S.~Bravyi and A.~Y. Kitaev,
Fermionic Quantum Computation,
\newblock Ann. Phys. (Amsterdam) {\bf 298}, 210 (2002), arXiv:0003137.

\bibitem{Backens:2019}
S.~Backens, A.~Shnirman, and Y.~Makhlin,
Jordan-Wigner transformations for tree structures,
\newblock Sci. Rep. {\bf 9}, 2598 ({2019}), arXiv:1810.02590.

\bibitem{Klco:2018zqz}
N.~Klco and M.~J. Savage,
Digitization of scalar fields for quantum computing,
\newblock Phys. Rev. A {\bf 99}, 052335 (2019), arXiv:1808.10378.

\bibitem{Byrnes:2005qx}
T.~Byrnes and Y.~Yamamoto,
Simulating lattice gauge theories on a quantum computer,
\newblock Phys. Rev. A {\bf 73}, 022328 (2006), arXiv:quant-ph/0510027.

\bibitem{Zhang:2018ufj}
J.~Zhang, J.~Unmuth-Yockey, J.~Zeiher, A.~Bazavov, S.-W.~Tsai, and Y.~Meurice,
Quantum Simulation of the Universal Features of the Polyakov Loop,
\newblock Phys. Rev. Lett. {\bf 121}, 223201 (2018), arXiv:1803.11166.

\bibitem{Unmuth-Yockey:2018xak}
J.~F. Unmuth-Yockey,
Gauge-invariant rotor Hamiltonian from dual variables of 3D $U(1)$ gauge theory,
\newblock Phys. Rev. D {\bf 99}, 074502 (2019), arXiv:1811.05884.

\bibitem{Alexandru:2019nsa}
A.~Alexandru, P.~F.~Bedaque, S.~Harmalkar, H.~Lamm, S.~Lawrence, and N.~C.~Warrington (NuQS Collaboration),
Gluon field digitization for quantum computers,
\newblock Phys. Rev. D {\bf 100}, 114501 (2019), arXiv:1906.11213.

\bibitem{Ji:2020kjk}
Y.~Ji, H.~Lamm, and S.~Zhu (NuQS Collaboration),
Gluon field digitization via group space decimation for quantum computers,
\newblock Phys. Rev. D {\bf 102}, 114513 (2020), arXiv:2005.14221.

\bibitem{Lamm:2019bik}
H.~Lamm, S.~Lawrence, and Y.~Yamauchi (NuQS Collaboration),
General methods for digital quantum simulation of gauge theories,
\newblock Phys. Rev. D {\bf 100}, 034518 (2019), arXiv:1903.08807.

\bibitem{Brower:2020huh}
R.~C. Brower, D.~Berenstein, and H.~Kawai,
Lattice Gauge Theory for a Quantum Computer,
\newblock Proc. Sci. {\bf LATTICE2019}, 112 (2020), arXiv:2002.10028.

\bibitem{Kreshchuk:2020dla}
M.~Kreshchuk, W.~M. Kirby, G.~Goldstein, H.~Beauchemin, and P.~J. Love,
Quantum simulation of quantum field theory in the light-front formulation,
\newblock Phys. Rev. A {\bf 105}, 032418 (2022), arXiv:2002.04016.

\bibitem{farhi2014quantum}
E.~Farhi, J.~Goldstone, and S.~Gutmann,
A Quantum Approximate Optimization Algorithm,
\newblock arXiv:1411.4028.

\bibitem{QAOA}
S.~Hadfield {\em et~al.},
From the Quantum Approximate Optimization Algorithm to a Quantum Alternating Operator Ansatz,
\newblock Algorithms {\bf 12}, 34 (2019), arXiv:1709.03489.

\bibitem{wiersema-PRXQuantum2020}
R.~Wiersema, C.~Zhou, Y.~de~Sereville, J.~F.~Carrasquilla, Y.~B.~Kim, and H.~Yuen,
Exploring Entanglement and Optimization within the Hamiltonian Variational Ansatz,
\newblock PRX Quantum {\bf 1}, 020319 (2020).

\bibitem{nakanishi_19}
K.~M. Nakanishi, K.~Mitarai, and K.~Fujii,
Subspace-search variational quantum eigensolver for excited states,
\newblock Phys. Rev. Res. {\bf 1}, 033062 (2019).

\bibitem{Pedernales:2014}
J.~S. Pedernales, R.~Di~Candia, I.~L. Egusquiza, J.~Casanova, and E.~Solano,
Efficient Quantum Algorithm for Computing $n$-time Correlation Functions,
\newblock Phys. Rev. Lett. {\bf 113}, 020505 (2014), arXiv:1401.2430.

\bibitem{nielsen_chuang_2010}
M.~A. Nielsen and I.~L. Chuang,
\newblock {\em Quantum Computation and Quantum Information: 10th Anniversary
  Edition} (Cambridge University Press, Cambridge, England, 2010).

\bibitem{Pfeffer:2023yhb}
P.~Pfeffer,
Multidimensional Quantum Fourier Transformation,
arXiv:2301.13835 

\bibitem{Luscher:1990ux}
M.~Luscher,
Two-particle states on a torus and their relation to the scattering matrix,
\newblock Nucl. Phys. {\bf B354}, 531 (1991).

\bibitem{Briceno:2017max}
R.~A. Briceno, J.~J. Dudek, and R.~D. Young,
Scattering processes and resonances from lattice QCD,
\newblock Rev. Mod. Phys. {\bf 90}, 025001 (2018), arXiv:1706.06223.

\bibitem{Andersen:2018mau}
C.~Andersen, J.~Bulava, B.~H\"orz, and C.~Morningstar,
The $I=1$ pion-pion scattering amplitude and timelike pion form factor from $N_f=2+1$ lattice QCD,
\newblock Nucl. Phys. {\bf B939}, 145 (2019), arXiv:1808.05007.

\bibitem{Nambu:1961tp}
Y.~Nambu and G.~Jona-Lasinio,
Dynamical Model of Elementary Particles Based on an Analogy with Superconductivity. I,
\newblock Phys. Rev. {\bf 122}, 345 (1961).

\bibitem{Nambu:1961fr}
Y.~Nambu and G.~Jona-Lasinio,
Dynamical Model of Elementary Particles Based on an Analogy with Superconductivity. II,
\newblock Phys. Rev. {\bf 124}, 246 (1961).

\bibitem{Gross:1974jv}
D.~J. Gross and A.~Neveu,
Dynamical symmetry breaking in asymptotically free field theories,
\newblock Phys. Rev. D {\bf 10}, 3235 (1974).

\bibitem{quspin}
P.~Weinberg and M.~Bukov,
QuSpin: a Python package for dynamics and exact diagonalisation of quantum many body systems part I: spin chains,
\newblock SciPost Phys. {\bf 2}, 003 (2017).

\bibitem{Steiger2018projectqopensource}
D.~S. Steiger, T.~H{\"{a}}ner, and M.~Troyer,
ProjectQ: an open source software framework for quantum computing,
\newblock {Quantum} {\bf 2}, 49 (2018).


\end{thebibliography}

\end{document}